\documentclass[journal abbreviation]{copernicus2}

\usepackage{color}
\usepackage[T1]{fontenc}

\begin{document}

\title{Inverse scattering problem in turbulent magnetic fluctuations
}

\author[1]{R. A. Treumann\thanks{Visiting the International Space Science Institute, Bern, Switzerland}}
\author[2]{W. Baumjohann}
\author[2]{Y. Narita}

\affil[1]{Department of Geophysics and Environmental Sciences, Munich University, Munich, Germany}
\affil[2]{Space Research Institute, Austrian Academy of Sciences, Graz, Austria}

\runningtitle{Inverse problem of magnetic fluctuations}

\runningauthor{R. A. Treumann, W. Baumjohann and Y. Narita
}

\correspondence{R. A.Treumann\\ (rudolf.treumann@geophysik.uni-muenchen.de)}


\firstpage{1}

\maketitle

{\subsection*{\bf{Abstract}} 
We apply a particular form of the inverse scattering theory to turbulent magnetic fluctuations in a plasma. In the present note we develop the theory, formulate the magnetic fluctuation problem in terms of its electrodynamic turbulent response function, and reduce it to the solution of a special form of the famous Gel$'$fand-Levitan-Marchenko equation of  quantum mechanical scattering theory. The latter applies to transmission and reflection in an active medium. {Theory of turbulent magnetic fluctuations} does not refer to such quantities. It requires a somewhat different formulation. {We  reduce the theory to the measurement of the low-frequency electromagnetic fluctuation spectrum, which is not the turbulent spectral energy density!} The inverse theory in this form enables obtaining information about the turbulent response function of the medium. The dynamic causes of the electromagnetic fluctuations are implicit to it. Thus it is of vital interest in low frequency magnetic turbulence. The theory is developed until presentation of the equations in applicable form to observations of turbulent electromagnetic fluctuations as input from measurements. Solution of the final integral equation should be done by standard numerical methods based on iteration.  We point on the possibility of treating power law fluctuation spectra as an example. Formulation of the problem to include observations of spectral power densities in turbulence is not attempted. This leads to severy mathematical problems and requires a reformulation of inverse scattering theory.  One particular aspect of the present inverse theory of turbulent fluctuations is that its structure naturally leads to spatial information which is obtained from the temporal information that is inherent to the obseration of time series. The Taylor assumption is not needed here. This is a consequence of Maxwells equations which couple space and time evolution. The inversion procedure takes advantage of a particular mapping from time to space domains. Though the theory is developed for homogeneous stationary non-flowing media, its extension to include flows, anisotropy, non-stationarity and the presence of spectral lines, i.e. plasma eigenmodes like present in the foreshock or the magnetosheath is obvious. }
\vspace{0.5cm}

\section{Introduction}
\label{intro}
As far as it concerns the fluctuations of the magnetic field, magnetic turbulence \citep{goldstein1995,biskamp2003,zhou2004,brown2015} is a branch of classical electrodynamics with the electromagnetic field described by Maxwell's equations. Its coupling to the dynamics of charged particles, ions and electrons, is contained in a set of separate dynamical equations. Depending on the spatial and temporal scales of the turbulence, these equations are subject to increasing simplifications. 

On the shortest scales $\ell< r_{ce}=v_e/\omega_{ce}$ shorter than the electron gyroradius $r_{ce}$ any turbulence is about purely electric/electrostatic as long as the plasma is not subject to self-magnetisation via excitation of either Weibel-like modes or nonlinear ion and electron holes, as also when spontaneous reconnection in electron-scale current filaments comes into play. The latter is believed to contribute an ultimate dissipation mechanism for collisionless turbulence \citep[cf., e.g.,][]{treumann2015}. The dynamic part of the turbulence is described by electrostatic kinetic equations and the talk goes of plasma turbulence. 

On longer scales $r_{ce}<\ell< r_{ci}=v_i/\omega_{ci}$ between the electron and ion gyroradii, electrons magnetise and thus contribute to magnetic turbulence. The magnetically active turbulent frequencies in this range are below the electron cyclotron frequency $\omega<\omega_{ce}$. Coupling of electrons to the nonmagnetic ions however connects the magnetic fluctuations with electrostatic ion-fluctuations as is, for instance, the case in the presence of kinetic Alfv\'en waves. In this range of scales electrons carry the magnetic field and also form the narrow turbulent current filaments. The system in this range is highly nonlinear and too complex for taking into account the plasma dynamics in generation of the turbulence. 

On the other hand, measurement of the magnetic fluctuations is comparably easy in plasma. One would thus like to infer about the turbulent plasma dynamics from the magnetic fluctuations alone, if possible. This is usually done from observation of the magnetic power spectra of turbulence and determination of the spectral index in several ranges of scales, from magnetohydrodynamic scales down into the dissipative range of scales of turbulence. This procedure mainly provides power law indices of the magnetic turbulence and  distinguishes between different spectral ranges and between inertial and dissipation scales while no information about the state of the plasma can be obtained.

In the following we attempt a different approach by formulating a so-called ``inverse problem" for the particular case of magnetic turbulence. This is possible when recognising that, as noted above, magnetic turbulence is in fact just a branch of classical electrodynamics. It can thus be formulated in terms of purely electrodynamic quantities with the dynamics implicitly included only. 

{In a first step of such an approach, we demonstrate, how the problem of magnetic fluctuations} can be reduced to the solution of an inverse problem, whose solution is, of course, nontrivial. We develop the theory until the formulation of the final integral equation whose input is the experimentally obtained field fluctuation spectrum. ({This is not the power spectral density usually used in turbulence and inferred by measurements. Instead, it is the full spectrum of electro-magnetic fluctuations that is on stake -- quite different from the magnetic power spectral densities used in ordinary low frequency turbulence!}) This integral equation will have to be solved for any given observed fluctuation spectrum. This reformulation of magnetic fluctuation theory might provide a new path {in the investigation of turbulence as it infers about the dynamics of the plasma which leads to the generation of turbulent fluctuations. In a subsequent step its relation to observed turbulent spectral power densities should be investigated.} Here we just develop the inverse magnetic {fluctuation} theory. A similar approach should, of course, also be possible for genuine kinetic plasma turbulent fluctuations including electron scales.

\section{Electro-magnetic turbulence equation}
{We restrict to purely magnetic turbulence, i.e. turbulent fluctuations $\delta\vec{B}$ in the magnetic field $\vec{B}$  caused by plasma motions on the scales under consideration. Such fluctuations of the magnetic field also include quasi-neutral fluctuations in the electric induction field $\delta\vec{E}$. Since electric charge fluctuations are absent $\delta\rho=e(\delta N_i-\delta N_e)=0$ yielding $N_i=N_e=N$, they are caused solely by plasma motions, pressure gradients etc. In this case and for the low frequencies expected the displacement current can be neglected. Maxwell's equations of electrodynamics apply in the Lorentz gauge which reduces to the radiation gauge.} The turbulent electric conduction current $\delta\vec{j}=eN(\delta\vec{V}_i-\delta\vec{V_e})$ in the plasma can then always in full generality be expressed by the product 
\begin{equation}
\delta\vec{j}=\hat{\vec{\sigma}}\cdot\delta\vec{E}
\end{equation}
of an appropriately defined conductivity tensor $\hat{\vec{\sigma}}$ and the fluctuating induction electric field $\delta\vec{E}$. 

Determination of the conductivity tensor depends on which dynamical reference model of the plasma is adopted. 
Herein lies the difficulty of relating the magnetic fluctuations as well as the magnetic spectral power density to the turbulent flow. In general, $\hat{\vec{\sigma}}$ is a functional of all plasma quantities. These, by reference to the most general form of the generalised Ohm's law, reduce to a functional of the electromagnetic field and, in the radiation gauge are reduced further to only one of the fluctuation  fields, the electric field $\delta\vec{E}$ or the magnetic field $\delta\vec{B}$. 

One may note that this \emph{ansatz} by no means implies linearity. Through all the plasma quantities and the electromagnetic fields, $\hat{\vec{\sigma}}(\vec{x},t)=\langle\hat{\vec{\sigma}}(\vec{x},t)\rangle+\delta\hat{\vec{\sigma}}(\vec{x},t)$ becomes a function of the spatial coordinates $\vec{x}$ and time $t$. {Also, in general an external magnetic field $\vec{B_0}$ may be present which penetrates the plasma and gives rise to an external current $\vec{j}_0$. Both may be composed of a generic external field and current or be mean-field quantities $\langle\vec{j}\rangle, \langle\vec{B}\rangle$ in the sense that they result from averaging $\langle\dots\rangle$ over the largest turbulent scales. The fluctuating current is defined as $\delta\vec{j}=\vec{j}-\langle\vec{j}\rangle-\vec{j}_0$. Correspondingly, the fluctuating magnetic field becomes $\delta\vec{B}=\vec{B}-\langle\vec{B}\rangle-\vec{B}_0$.} 

{In the following we deal with fluctuations only and for convenience drop the prefix $\delta$ on all quantities. External and mean fields do not vary on the scale of the fluctuations. If needed, they can be combined into one mean/external field quantity $\langle\vec{B}\rangle+\vec{B}_0\longrightarrow\vec{B}_0$.} 

\subsection{Fluctuations of vector potential}
With these preliminaries in mind Maxwell's equations for the fluctuating fields {in the absence of charge density fluctuations $\rho=0$} reduce to the following set of equations
\begin{eqnarray}
\nabla^2\vec{E}(\vec{x},t)~~&=&~\mu_0\partial_t\vec{j}(\vec{x},t)\\[0.5ex]
\nabla\times\vec{A}(\vec{x},t)&=&~\vec{B}(\vec{x},t),\quad\partial_t\vec{A}(\vec{x},t)=-\vec{E}(\vec{x},t)
\end{eqnarray}
All plasma dynamics in this representation is absorbed into the conductivity tensor. Since this dynamics is an internal plasma property, causality requires that the plasma response to any disturbance in the electromagnetic field, which gives rise to the conductivity tensor, depends only on the duration between the past time $t'$ and the time $t$ of observation. Thus $\hat{\vec{\sigma}}$ is a function of time difference $t-t'$, and the fluctuating current becomes a time-convolution integral 
\begin{equation}
\vec{j}(\vec{x},t)=\int_{-\infty}^t\!\!\!\!\mathrm{d}t'\hat{\vec{\sigma}}(t-t',\vec{x})\cdot\vec{E}(\vec{x},t')
\end{equation}
Fourier transformation of the above equations with respect to time then yields
\begin{eqnarray}
\nabla^2\vec{E}(\vec{x},\omega)~~&=&~-i\omega\mu_0\hat{\vec{\sigma}}(\vec{x},\omega)\cdot\vec{E}(\vec{x},\omega)\\[0.5ex]
\nabla\times\vec{A}(\vec{x},\omega)&=&~\vec{B}(\vec{x},\omega),\quad i\omega\vec{A}(\vec{x},\omega)=\vec{E}(\vec{x},\omega)
\end{eqnarray}
Hence, solution of the first equation for the turbulent electric field completely determines the problem for any set of given intial and boundary conditions. Solving for the Fourier transform of the electric field  determines the vector potential and from it the magnetic field, both as functions of the frequency of the turbulent fluctuations. From that point of view is is convenient to work with the vector potential and replace the turbulent electric field:
\begin{equation}\label{eq-eqa}
\nabla^2\vec{A}(\vec{x},\omega)=-i\omega\mu_0\hat{\vec{\sigma}}(\vec{x},\omega)\cdot\vec{A}(\vec{x},\omega)
\end{equation}
What concerns the boundary conditions in a nominally infinite medium like an unbounded plasma one just requires that the fields are analytical at $\pm\infty$ in space. Otherwise, if boundaries are given, one requires continuity of the tangential components of $A$ and its spatial derivative at the boundaries. 

The difficulty in solving the above equation for a turbulent plasma lies in the fact that the generalised conductivity tensor $\vec{\hat\sigma}$ is barely known. Prescription of some analytical form depends on the chosen model. For this reason a direct application to turbulence is illusive even though electrodynamics seems to provide a simple approach to turbulence. This is in fact not the case. One therefore may ask what can be inferred about the unknown turbulent conductivity if turbulence is observed and measured?

Before coming to this question, let us rewrite the last above equation in different form by introducing the dielectric response tensor of the turbulent plasma
\begin{equation}
\hat{\vec{\epsilon}}(\omega,\vec{x})=\frac{\hat{i\vec{\sigma}}(\omega,\vec{x})}{\epsilon_0\omega}=\langle\hat{\vec{\epsilon}}(\omega,\vec{x})\rangle+\delta\hat{\vec{\epsilon}}(\omega,\vec{x})
\end{equation}
In terms of this quantity the equation for the turbulent field $\vec{A}(\omega,\vec{x})$ becomes {(after subtraction of the fluctuation-averaged equation and dropping the mean-field term $\langle\delta\hat{\vec\epsilon}\cdot\delta\vec{A}\rangle$, which on the fluctuation scale contributes a constant)}  
\begin{equation}\label{eq-aone}
\nabla^2\vec{A}(\vec{x},\omega)=\left(\frac{i\omega}{c}\right)^{\!\!2}\: \hat{\vec{\epsilon}}(\vec{x},\omega)\cdot\vec{A}(\vec{x},\omega)
\end{equation}
The factor $i\omega/c=k_0$ is a transverse ``turbulent wavenumber'' that through the frequency $i\omega=\vec{k}_0\cdot \delta\vec{v}$ is related to the turbulent velocity $\delta\vec{v}$, and $\hat{\vec{\epsilon}}=\vec{\epsilon}^T$ is the transverse  turbulent response function of the turbulent plasma which describes the low-frequency purely electromagnetic fluctuations. Note that there are no charge fluctuations and therefore no fluctuations of electrostatic potential. With its help we can write 
\begin{equation}
\nabla^2 \vec{A}_\omega(\vec{x})=k_0^2\vec{\epsilon}^T_\omega(\vec{x})\cdot\vec{A}_\omega(\vec{x})
\end{equation}
where the dependence on frequency has been absorbed into the index. In any measurement of turbulence the vector potential can be considered as known, i.e. given by the measurements. The completely unknown quantity here is the transverse response function. We therefore wish to find a relation that expresses the response function -- or otherwise the conductivity -- through the observations which to some extent are buried in $\vec{A}_\omega$.

{One remark on the fluctuation of the dielectric response tensor (suppressing the $\delta$-sign) is in place here. In turbulence, $\hat{\vec{\epsilon}}(\omega,\vec{x}, |\vec{A}_\omega|^2)$ depends explicitly on the vector potential $\vec{A}_\omega$ respectively the fluctuating electric field $\vec{E}_\omega$. This dependence is \emph{nonlinear}, implying that the fluctuations cause a turbulent modification of the dielectric response. For this reason the response depends on the squared magnitude $\vec{A}_\omega$. Expansion yields
\begin{eqnarray}
\hat{\vec{\epsilon}}\big(|\vec{A}_\omega|^2\big)&=&\hat{\vec{\epsilon}}_m(|\vec{A}_\omega|^2_m)+\\ 
&+&\frac{\partial\hat{\vec{\epsilon}}(|\vec{A}_\omega|^2)}{\partial |\vec{A}_\omega|^2}\Big(|\vec{A}_\omega|^2-|\vec{A}_\omega|^2_m\Big)+\cdots
\end{eqnarray}
The index $m$ indicates that the first term on the right is taken at  $|\vec{A}_\omega|^2_m$, the value  where the turbulent fluctuations exert an extremal effect on the response function. Hence $\partial\hat{\vec{\epsilon}}(|\vec{A}_\omega|^2)/\partial|\vec{A}_\omega|^2 =0$ and the term linear in $|\vec{A}|_\omega^2$ in the expansion disappears which implies interpreting $\vec{A}_\omega \approx \vec{A}_{\omega m}$. Retaining this term would transform Eq. (\ref{eq-aone}) into third-order nonlinear Schrödinger-like form. This is avoided by reference to the maximum nonlinear effect. The second order derivative is negative. Its inclusion corresponds to a term of fourth-order in the vector potential in the induction equation 
\begin{eqnarray}
\nabla^2 \vec{A}_\omega&=&k_0^2\Big[\vec{\epsilon}^T_{m\omega}(\vec{x},|\vec{A}_\omega|^2_m)+\\
&+&{\frac{1}{2}}\frac{\partial^2\vec{\epsilon}^T_{m\omega}(\vec{x},|\vec{A}_\omega|^2)}{\partial(|\vec{A}_\omega|^2)^{2}}\Big(|\vec{A}_\omega|^2-|\vec{A}_\omega|^2_m\Big)^{\!\!2}\:\bigg]\cdot\vec{A}_\omega\nonumber
\end{eqnarray}
This equation would  generate a relation to Ginsburg-Landau theory respectively a fourth-order potential allowing for phase transition and symmetry breaking effects which, in a forward theory, may become of interest. Our interest here is however on the inverse problem. We therefore below restrict to the investigation of the (in the above spirit) maximum nonlinear effect only and neglect both the first and the second-order derivative contributions.}

\subsection{Magnetic field fluctuations}
{So far we formulated the fluctuation problem in terms of the fluctuating induction electric respectively vector potential fields.} Alternatively, the entire problem can also be given a formulation purely in terms of the magnetic field. For this we define the inverse transverse response tensor $\vec{\eta}_\omega^T(\vec{x})=\left[\vec{\epsilon}_\omega^{T}(\vec{x})\right]^{-1}$. With $\vec{E}(\omega,\vec{x})=i\omega\vec{A}(\omega,\vec{x})$, multiplying Amp\`ere's law by $\vec{\eta}^T_\omega$ and making straightforwardly use of $\vec{B}=\nabla\times\vec{A}$ one obtains 
\begin{equation}
\nabla\times\vec{\eta}^T_\omega(\vec{x})\cdot\nabla\times\vec{B}(\omega,\vec{x})=k_0^2\vec{B}(\omega,\vec{x})
\end{equation}
This equation contains only the magnetic field in explicit form. It looks somewhat more complicated than Eq. (\ref{eq-eqa}), because it contains the spatial derivative of the inverse response tensor $\vec{\eta}_\omega^T$. It -- apparently -- has the advantage that one needs not to refer to electric field observations. Its one-dimensional version applicable to the fluctuation problem is
\begin{equation}\label{eq-newind-b}
\partial_x\left[\eta_\omega^T(x)\partial_x B_\omega(x)\right]=k_0^2B_\omega(x)
\end{equation}
which follows when $\vec{B}$ has only the component $B_z(x)$. We will show later that this apparent advantage is lost in reality because the complete electromagnetic information is available only if both the electric and magnetic field fluctuations are available. This becomes crucial in the inverse problem which can be solved only if the {experimental electromagnetic} field data are complete.

\section{Relation to observations}
Observations are frequently just one-dimensional. Magnetic fluctuations are strictly transverse with solenoidal fields $\nabla\cdot\vec{B}=0$. In the presence of an exernal field $\vec{B}_0$ -- or equivalently a non-vanishing mean field $\langle\vec{B}\rangle$ which may be included into the external field -- {turbulent fields are usually oblique. They can be divided into parallel to $\vec{B}_0$ and perpendicular fluctuations. Parallel fluctuations propagate perpendicular, perpendicular propagate parallel to the field}. 

In developing the theory, let us for simplicity restrict to the case of {perpendicular magnetic fluctuations propagating along the mean field.} We assume that {the external/mean field is in direction $\vec{B}_0=B_0 \vec{e}_x$, and the fluctuations are in the direction $\vec{B}=B_z\vec{e}_z$. Thus all wavenumbers of the turbulent fluctuations are parallel to the external/mean field.} Then $B_z=\partial_xA_y(x)$, and the fundamental equation to solve becomes
\begin{equation}\label{eq-newind1}
A_{y\omega}''(x)=k_0^2\epsilon^T_\omega(x)A_{y\omega}(x)
\end{equation}
Here $'$ denotes differentiation with respect to the argument, in the present case $x$. From now on we drop the spatial indices. 

The observations refer to the fluctuations of the magnetic field in time from which the temporal spectrum is  subsequently obtained. The determination of the spatial spectrum is difficult. It requires the simultaneous measurement of the magnetic field over an extended area in various locations, for various spacings and many frequencies. In near-Earth space where this is mostly done -- a well-known example is the solar wind -- frequency spectra have been obtained in multitude. However, so long multi-spacecraft missions are rare. Only very few direct measurements of spatial spectra have become available in a restricted range of wavenumbers. 

Measured time series of the magnetic fluctuation field $B(t,x_0)$ at particular fixed observation site $x_0$ transform directly into their Fourier equivalent 
\begin{equation}
B_\omega(x_0)=\lim\limits_{x\to x_0}\int_{-\infty}^\infty \mathrm{d}t\ B(t,x)\:\mathrm{e}^{i\omega t}
\end{equation}
or in terms of the vector potential 
\begin{equation}
B_\omega(x_0)=\lim\limits_{x\to x_0} \partial_x A_\omega(x)
\end{equation}
Such measurements  are subsequently used to obtain spectral energy densities of the magnetic turbulence. It is this quantity to that turbulence theory refers. 

No direct spatial dependence of $A(t,x)$ is provided by the data. Neither is the equivalent fluctuating electric field $E_\omega(x)=i\omega A_\omega(x)$ known because of the experimental difficulties involved in measuring the low frequency turbulent electric field. This is in contrast to the high frequency electric wave fields. The latter are easily obtained with the help of antennas or probes. 

Techniques of measuring the turbulent electric field by following the electron gyration orbit have been developed and are sometimes available. However no application to turbulence has been attempted because these techniques are highly complex. It would be of great advantage if they would be transformed into measuring devices for the direct determination of the turbulent vector potential. 

To make them efficient they should be applied to ion gyrations. Ion orbits are substantially more stable than electron orbits. They therefore could map the low frequency turbulent electric field in the solar wind or elsewhere. We will show below that availability of both magnetic and electric time series provides a device for investigation of turbulence.  

\subsection{Taylor range}
Temporal spectra are frequently transformed into spatial spectra under the assumption of Taylor's hypothesis. This hypothesis implies that for sufficiently low frequencies and sufficiently fast plasma streams the turbulent eddies and fluctuations are swept across the spacecraft, Galilei-transforming wavenumbers $\vec{k}$ into frequency space $\omega$ according to
\begin{equation}
\omega=\omega'-\vec{k}\cdot\vec{V}_0
\end{equation}
$\omega'$ is the frequency in the moving frame. If  $\omega'\ll\vec{k\cdot V}_0$ the frequency $\omega$ maps into wavenumber $\vec{k}$ along the streaming velocity $\vec{V}_0$. Dividing by $k$  this simply means that the phase velocity must be much less than the speed of streaming. 

Spectra propagating perpendicular to the streaming are not affected. Their wavenumber spectra cannot be obtained in this simple way. These conclusions are well known. Checking for the solar wind one immediately finds that all low frequency waves obey the Taylor hypothesis if propagating along the solar wind. Their nominal speed is the Alfv\'en speed $\omega/k\sim V_A$. Since the solar wind is super-Alfv\'enic with $4\lesssim V_0/V_A\lesssim 10$ the parallel $k$-spectrum is recovered from the frequency spectrum. In the electron-whistler and kinetic-Alfv\'en range the mapping becomes problematic. Electron whistler speeds are of the order of $\omega/k\sim 40\:V_A$ making any Galilei-mapping obsolete. 

On the other hand, there is no reason to assume that fluctuations and turbulent spectra consist of eigenmodes. At the contrary, they do not show indication of line spectra which would correspond to particular plasma wave modes. Instead they represent fluctuations transformed into wavenumber and frequency space which may not have anything to do with eigenmodes even when resulting originally from the injection of energy at injection wavenumbers by some plasma eigenmode spectrum. Turbulent cascading is highly nonlinear and obscures any plasma modes which might be present inside its range of wavenumbers. For this reason the Taylor mapping may well hold in all cases as long as the ratio $\omega/k<V_0$ is sufficiently small and $k$ is parallel to the flow.   

The Taylor hypothesis becomes problematic in anisotropic turbulence \citep{horbury2012,wicks2012}.
In the range of its validity one replaces the $x$-dependence with a time-dependence via $x=V_0\tau$. This is, however, no advantage because the plasma response function is not known, and therefore measuring the magnetic fluctuations does not provide information about the fluctuations of the induction field. The relation between them is given simply by the streaming velocity $V_0$: 
\begin{equation}
E_k(x_0)= \omega B_k(x_0)/k= V_0B_k(x_0)
\end{equation}
an expression which is of no interest as it leaves our knowledge about the electromagnetic fluctuation field incomplete. Therefore we must look for some other way than the Taylor hypothesis of making use of field measurements.

\subsection{Correlation function of magnetic fluctuation field}
By the same token it is obvious that the sole measurement of magnetic power spectra provides incomplete information only about magnetic turbulence. 

This can be shown more explicitly. The turbulent magnetic power spectrum is the only solid measurement available when no electric field observations are available und no velocity fluctuations can be obtained. The latter could, as well in principle, be constructed from measurements of the distribution function if only resolution in velocity/momentum and time would suffice. In that case the turbulence problem would have to be formulated differently by reference to the dynamic equations or, more directly, to the kinetic  equations.

In the availability of only magnetic fluctuations, one forms the average time correlation function of the magnetic field at location $x_0$, the measured quantity, averaged over the observation period $T$,
\begin{eqnarray}
&&\Big\langle B(t-\tau,x_0)B(t,x_0)\Big\rangle_{\!\!(T)} = \nonumber\\
&&\qquad\qquad=\lim_{T\to\infty}\frac{1}{2\pi T}\int_{-{T/2}}^{T/2}\mathrm{d}\omega \big|B_\omega(x_0)\big|^2\mathrm{e}^{-i\omega\tau}
\end{eqnarray}
with index $T$ indicating that the average is with respect to the observational time interval $-\frac{1}{2}T\ \leq\ t\ \leq\ \frac{1}{2}T$. Its limit $\tau\to 0$ is the energy density of the turbulence, and $|B_\omega|^2\mathrm{d}\omega$ is the spectral energy contained in the frequency interval $\mathrm{d}\omega$ at location $x_0$. This expression is a Fourier transform from which we have that 
\begin{equation}\label{eq-b2}
\big|B_\omega(x_0)\big|^2 =\int\mathrm{d}\tau\Big\langle B(t,x_0)B(t-\tau,x_0)\Big\rangle_{\!\!\!\!({T})}\mathrm{e}^{i\omega\tau}
\end{equation}
is the Fourier-transformed correlation function at location $x_0$ with respect to the time interval $T$ of observation. Since the right hand side has been measured, the Fourier transform can be obtained, and if the limit $T\to\infty$ can be performed, the left-hand side is a known quantity. It forms the usual experimental background of any investigation of magnetic turbulence in a collisionless plasma like the solar wind. If one would know the dispersion relation $\omega(k)=\omega_k$ of the turbulence, one could directly infer the spectral energy density in wave number space by the transformation $\mathrm{d}\omega\to(\partial\omega_k/\partial k)\mathrm{d}k$. Turbulence does not possess any well defined dispersion relation, however.

Of course, assuming that $x_0$ would be a continuous variable then, as done in spectral theory, one can average over the spatial coordinate and expand as well with respect to space. The spectral density of the correlation function with respect to time and space then becomes
\begin{equation}
\big|B_{\omega k}\big|^2\!\!\! =\!\!\! \int\mathrm{d}\tau\mathrm{d}\xi\Big\langle\!\! B(t,x)B(t-\tau,x-\xi)\!\!\Big\rangle_{\!\!\!\!\mathit{\!(\!TX\!)}}\mathrm{e}^{i(\omega\tau-k\xi)}
\end{equation}
where the average is taken over the space-time interval $[(T),(X)]$. From spectral theory it is known that the correct spectral energy density in wavenumber space is obtained from
\begin{equation}
\big|B_k\big|^2\sim\int\mathrm{d}\omega\:\big|B_{\omega k}\big|^2
\end{equation}
This well-known relation replaces Taylor's hypothesis. Unfortunately, $\big|B_{\omega,k}\big|^2$ is barely known from observations in space. What is known is just $\big|B_{\omega}(x_0)\big|^2$, the spectral energy density as function of frequency at the location of observation $x_0$. Hence, important information is missing. For this reason one is forced to grab for reasonable models of the response function in order to reproduce the observed spectra. This means solving the forward problem of turbulence.

\subsection{Dissipative convolution function}
Another promising measurable quantity that can be constructed from observations is the dissipative convolution function 
\begin{equation}
c_\omega(x-x_0)= -\frac{E_\omega(x-x_0)}{E'_\omega(x-x_0)}=\frac{iE_\omega(x-x_0)}{\omega B_\omega(x-x_0)}
\end{equation}
Its construction requires independent measurements of the two electromagnetic time series $B(t),\: E(t)$. If this is possible, then the full information about the electromagnetic fluctuation field is available. 

In a magnetically turbulent medium, the convolution function $c_\omega(x)$ contains the full nonlinear low-frequency electromagnetic response of the plasma. It excludes purely electrostatic fluctuations which enter the kinetic high frequency short-scale turbulence. It also excludes the range of free-space electromagnetic waves. Its disadvantage is that as long as the two time-series are not available its use is inhibited. 

 $c_\omega(x)$ has the dimension of  a length. Multiplying $|c_\omega|^2=c_\omega c^*_\omega$ by $\omega\mu_0$ becomes an equivalent resistivity and thus directly measures the total turbulent dissipation. It is but an apparent dissipation whose relation to the real dissipation must be established in some way.

To demonstrate that $c_\omega$ is the convolution of the turbulent electromagnetic field fluctuations, one observes that the three quantities in the last expression are all Fourier transforms with respect to time. Analyticity is warranted by the fact that the fluctuating electric field
\begin{equation}
E_\omega(x-x_0)=-c_\omega(x-x_0)E'_\omega(x-x_0)
\end{equation}
is the product of two frequency-time Fourier transforms and hence is the Fourier transform of the convolution integral
\begin{equation}
E(t,x)= \int_0^t\mathrm{d}\tau\ c(t-\tau,x)\frac{\partial}{\partial\tau} B(\tau,x)
\end{equation}
which by Faraday's law includes the magnetic field. The folding is provided by the analytic function $c(x,t)$ obtained from observations. In the solar wind and magnetospheric tail  $c(x,t)$ is usually not available because either its dependence on space $x$ cannot be determined or no electric fields are measured.

In homogeneous turbulence all quantities depend on spatial differences $x-x_0$ only, and it  makes no difference at which location the measurements have been performed. This means that $c_\omega(x)$ also implies a spatial folding. We have
\begin{equation}
E_{\omega\kappa}=i\omega c_{\omega\kappa}B_{\omega\kappa}
\end{equation}
an expression that corresponds to the space-time convolution integral
\begin{equation}
E(t,x)= \int\limits_0^{t}\!\!\!\!\!\!\!\int\limits_0^x \mathrm{d}\tau\: \mathrm{d}\chi\ c(t-\tau,x-\chi)\,\frac{\partial}{\partial\tau} B(\tau,\chi)
\end{equation}

In the following we show that the dissipative convolution function $c_\omega$ provides the key to the information about the turbulent dielectric function, the crucial quantity that contains the dynamic effect on the generation of magnetic fluctuations. Use of the convolution function allows for a consistent formulation of the inverse problem of magnetic fluctuations in magnetic turbulence. This differs from the conventional use of spectral power densities of the turbulent magnetic field to which turbulence theory usually refers. Power spectral densities provide spectral indices which are interpreted as either indications of inertial ranges in turbulence, injection or dissipation regimes. These must be interpreted by invoking models of turbulent energy transfer and dissipation. 

The goal of our present approach is different. We try to use the electromagnetic fluctuation spectra as combined in the dissipative convolution function, a measurable quantity, in order to reconstruct the electromagnetic response of the plasma. This is not impossible because we are in possession of the evolution equation of the fluctuations. The idea is to invert this equation and to infer about the plasma response. This leads us to the inverse fluctuation problem.

\section{Inverse problem of turbulent fluctuations}
{We assume that the convolution function is available. It stores the observational information about the electromagnetic fluctuations. Is there a way of reconstructing the generalised dielectric response function $\epsilon^T_\omega$ from it and infer about the physical nature of the turbulence?}  Answering this question implies the solution of the inverse problem of one of the two fundamental equations for the electromagnetic fluctuation field. 

{The inverse problem theory, known as the inverse scattering problem, was formulated by \citet{gelfand1951} in an attempt to reconstruct the scattering potential field from the quantum mechanical scattering matrix. In order to apply it to our problem, the induction equation for the fluctuation has to be brought into Schr\"odinger-like form. A similar problem has been pioneered in completely different context by \citet{weidelt1972} when dealing with the geomagnetic induction problem \citep{bailey1970,parker1980}. Its feasibility was demonstrated by \citet{weidelt2005} for a thin sheet conductivity model of a layered Earth. This problem is similar to the turbulent fluctuation problem with, however, a completely different form of data input. It thus requires its reformulation for our needs.} 

In a first step we transform the induction equation for the fluctuating vector potential into a Schr\"odinger equation.  
The necessary transformations are 
\begin{equation}
\xi :\ =\ k_0, \quad  \zeta_\xi(x):\ = \int_0^x \mathrm{d}y\sqrt{\epsilon_\xi^T(y)} 
\end{equation}
for the variables, and for the functions
\begin{eqnarray}
&& u_\xi(\zeta):\ =\ \Big[\epsilon^T_\xi(\zeta)\Big]^\frac{1}{4}\\[0.5ex]
&& f_\xi(\zeta):\ =\ u_\xi(\zeta){A_\xi(\zeta)} 
\end{eqnarray}
In these expressions the normalised frequency $\xi$ appears as a parameter only. It has the dimension of a reciprocal length or wavenumber $k_0$. For real frequency $\omega$, the imaginary phase of $\xi$ is $\pm i\pi/4$.  {One notes that the new spatial coordinate $\zeta$ is stretched by the turbulent dielectric}, a dimensionless quantity in our units containing the turbulent response of the plasma. 

Let $'$ indicate differentiation with respect to the argument, i.e. the genuine spatial coordinate $\zeta$. The basic electrodynamic equation for the vector potential then transforms into the standard Sturm-Liouville (stationary Schr\"odinger equation) form
\begin{equation}\label{eq-newind}
f''_\xi(\zeta)\:=\: \Big[\xi^2+V_\xi(\zeta)\Big]f_\xi(\zeta)
\end{equation}
This is shown in the Appendix by straightforward though quite tedious algebra. The fundamentally changed role of $\xi$ becomes clear from this familiar form where $\xi$ appears as kind of an ``eigenvalue". 

The ``equivalent potential'' function $V(\zeta,\xi)$ is defined uniquely through  $\epsilon^T_\xi(\zeta)$ respectively $u_\xi(\zeta)$, the stretching factor,  as
\begin{equation}
V_\xi(\zeta) \equiv u''_\xi(\zeta)\big/ u_\xi(\zeta)
\end{equation}
It is this function which is considered as unknown in our inversion problem. It is to be determined by solving the inverse problem. 

{One may note that the above stationary Schrödinger equation depends only on space. Frequency appears as an eigenvalue parameter. Its solution will depend on full space while observations exist just in one space point $x_0$.} 

This requires concluding about spatial variation of the fluctuations from observation in only one point. {As shown below, this becomes indeed possible due to the  transformation of frequency $\omega$ into wave number $\xi\sim k_0$. In this way $\xi$ becomes the conjugate to the spatial variable $\zeta$.} 

The standard Schr\"odinger form has a number of advantages. 
Its solution has been developed to very high standards. This provides a starting point for investigation of turbulence theory. {However, our  is only on the inversion of the above equation by making use of its known solutions.} 

The transformation of Eq. (\ref{eq-newind1}) can also be applied to the equation for the fluctuating magnetic field ({\ref{eq-newind-b}). This is not immediately obvious. Algebraically the calculation becomes quite involved while leading to an equation of the same kind as (\ref{eq-newind})
\begin{equation}\label{eq-newind-c}
\bar f_\xi''(\zeta)\:=\: \Big[\xi^2+\bar V_\xi(\zeta)\Big]\bar f_\xi(\zeta)
\end{equation}
where the ``equivalent potential" in the Schr\"odinger equation now turns out negative
\begin{equation}
\bar V_\xi(\zeta) \equiv - \bar u''_\xi(\zeta)\big/ \bar u_\xi(\zeta)
\end{equation}
Here the following definitions hold
\begin{equation}
\xi :\ =\ k_0, \quad  \zeta_\xi(x):\ = \int_0^x \mathrm{d}y\Big/\sqrt{\eta_\xi^T(y)} 
\end{equation}
for the variables, and for the functions
\begin{eqnarray}
&& \bar u_\xi(\zeta):\ =\ \Big[\eta^T_\xi(\zeta)\Big]^{-\frac{1}{4}}\\[0.5ex]
&& \bar f_\xi(\zeta):\ =\ \bar u_\xi(\zeta){B_\xi(\zeta)} 
\end{eqnarray}
These expressions are formally identical with those used in Eq. (\ref{eq-newind}) and, consequently, so is the formalism of solving the inverse problem. {The apparent advantage of dealing with magnetic fluctuations is, however, misleading. Reference to the electric induction field cannot be avoided for the reason of completeness}  

The upshot of the above relations is that the turbulent response function $\epsilon_\xi^T(\zeta)$ respectively its inverse, the turbulent dissipative response $\eta_\xi^T(\zeta)$ can, in principle, be determined if the solutions $f_\xi(\zeta)$ or $\bar f_\xi(\zeta)$ of the Schr\"odinger equation for the fluctuation problem can be constructed. {In this case it should be possible to invert the Schr\"odinger equation when understanding it as an equation for the equivalent potential $V_\xi(\zeta)$ (or $\bar{V}_\xi(\zeta)$) with given ``wave function" $f_\xi(\zeta)$ (or $\bar f_\xi(\zeta)$). The condition is that this solution can be expressed through the observations.} 

This problem is the \emph{inverse fluctuation problem of magnetic turbulence}. The formal problem of inverting the Schr\"odinger equation {was solved half a century ago \citep[cf.,][]{gelfand1951,marchenko2011}. It found wide application in quantum scattering theory, astrophysics,  geophysical exploration \citep[][]{weidelt1972,parker1980,carroll1981} as well as in the celebrated inverse scattering theory of the solution of the Korteweg-de Vries \citep[invented by][]{gardner1967} and other nonlinear equations.} Hence, attempting its application in turbulent fluctuation theory is by far not academic only. Its difficulty is encountered in relating the formal theory to observations. 

\subsection{Transformed convolution function}
In inverse scattering theory the available observations are the reflection and transmission coefficients of radiation \emph{passing} a passive medium. These coefficients are the ratios of the wave amplitudes of incoming from $-\infty$ and outgoing to $\pm\infty$ (including scattered/reflected) waves. In  turbulence and fluctuation theory these quantities are not available. There is no incident wave (if not explicitly inserted as a large-scale single wave source, which could in principle be included  -- transforming the stationary theory into an initial value problem of the evolution of turbulence). {Moreover, the turbulent medium is active in the sense that scattering, transmission, reflection and dissipation of the propagating fluctuations are intrinsic properties. One thus has to make use of other specific relations between data and the solutions of the Schr\"odinger equation.} 

One such possibility is reference to  the convolution function
$c_\omega(x-x_0)$. This is most simply expressed in terms of the vector potential
\begin{equation}
c_\omega(x)=-{A_\omega(x)\big/A'_\omega(x)}
\end{equation}
where, for simplicity, we have  put $x_0=0$.  Again $'$ indicates differentiation with respect to the argument $x$. As before, the dimension of $c_\omega(x)$ is that of a length or inverse wavenumber $c_\omega\sim k_0^{-1}$. {Expressing it through the vector potential will enable us to relate it to the two linearly independent solutions} of the equation for the vector potential $A_\omega(x)$. It has the properties
\begin{equation}
c_\omega^*(x)=c_{-\omega}(x) 
\end{equation}
which transforms into
\begin{equation}
c_\omega(x)=c_\xi(x), ~~ c_{\xi^*}(\zeta) = c^*_\xi(\zeta)
\end{equation}
Let the two independent appropriately normalised solutions of (\ref{eq-newind1}) for $A_\omega(x)$ be 
\begin{equation}
w_\omega^\pm\equiv A_\omega^\pm/A_\omega(0),\:w'_\omega\equiv A'_\omega/A_\omega(0)
\end{equation}
with ``initial conditions'' 
\begin{equation}
w_\omega^{+}(0)=1,\:w'^{-}_\omega(0)=k_0 =\xi, ~~w'^{+}_\omega(0)=w_\omega^{-}(0)=0
\end{equation}
at the location of observation $x=x_0=0$. Normalisation does not imply that the amplitude of $A_\omega$ has an extremum at $x_0$. If $x_0$ is the point of measurement then this is just the value of the turbulent vector potential at this location. On the other hand, $x_0$ can also be taken as location of the distant boundary of the turbulent volume, {in which case the vector potential is normalised to its value at this boundary. One then requires some boundary condition warranting continuity of the fields, i.e. continuity of $A_\omega$ and its derivative $A'_\omega$. If the medium streams with injection speed $V_0$ at this boundary, Galilei transformation implies $\omega=\omega'-kV_0$, where $k\neq k_0$ is the turbulent wave number. All these cases can in principle be included. What concerns the turbulent wavenumber $k$, so it satisfies some (highly nonlinear) unkown turbulent dispersion relation which is not obtained from  any  linear analysis. }

The solutions $w_\omega^\pm(x)$ are linearly independent. They satisfy the Wronskian determinant  
\begin{equation}
w_\omega^+(x)w'^-_\omega(x)-w'^+_\omega(x)w_\omega^-(x)=k_0\ \longmapsto\ \xi 
\end{equation}
This is easily checked because it holds for all (complex) $x$ and thus also for $x=x_0=0$. This condition can be exploited in relating the solution to observations. It can be used to eliminate the arbitrariness of the coeffcients of the full solution which is a linear combination of the two linearly independent solutions, implying that the total solution $A_\omega(x)$ can be written
\begin{equation}
A_\omega(x)=A_\omega(0)\left[w^+_\omega(x)-w_\omega^-(x)/c_\omega(0)\right]
\end{equation}
{This form follows from reference to the convolution function $c_\omega$, which is composed of the two independent solutions. At $x=0$ the boundary conditions of continuity imply that it has the finite value}
\begin{equation}
c_\omega(0)=-A_\omega(0)/A'_\omega(0) = -w_\omega^+(0)/w'^-_\omega(0) =-\xi^{-1}
\end{equation}
The last equation suggests that generally
\begin{equation}
c_\omega(x)= -w_\omega^+(x)/w'^-_\omega(x)
\end{equation}
Making use of the mappings one obtains
\begin{eqnarray}\label{eq-convc}
c_\omega(x)\ \longmapsto\ c_\xi(\zeta)&=&\frac{f_\xi^+(\zeta)}{f_\xi^-(\zeta) u'/u -f'^-_\xi(\zeta)}\\
c_\xi(0)&=&-f_\xi^+(0)/f'^-_\xi(0)
\end{eqnarray}
In the last experession $'$ means of course differentiation with respect to the argument, i.e. with respect to $\zeta$. 

The left-hand sides of these expressions are considered to be known from the data. They consist solely of data, while the right-hand sides are obtained from the linearly independent solutions $f^\pm_\xi(\zeta)$ of the Schr\"odinger equation 
\begin{equation}\label{eq-fxipm}
f_\xi(\zeta)=f_\xi^+-f_\xi^-/c_\xi(0)
\end{equation}
as representation of the full solution of the Schr\"odinger equation (\ref{eq-newind}). So far the solutions $f_\xi^\pm$ are unknown. They have to be determined as functions of the data, with the latter contained in $c_\xi(\zeta)$ or some derivative of it. 

{We note in passing that a similar procedure can be applied to the solutions of Eq. (\ref{eq-newind-c}) for the magnetic field fluctuations. However, in that case the  relation between the solutions of Schr\"odinger's equation and the convolution function become more involved}. 

It may be of some interest to ask for the relation to the observed magnetic spectrum. We have $B_\omega(x)=\partial_xA_\omega(x)$, and from the above transformations $\partial_x=u^{2}\partial_\zeta$. Therefore
\begin{eqnarray}\label{eq-rel-b-c}
B_\omega(x)\ &\longmapsto&\ u^2_\xi(\zeta)A'_\xi(\zeta) \\[0.5ex]
& \longrightarrow& u_\xi(\zeta)\Big[f'_\xi(\zeta)-\frac{u'_\xi(\zeta)}{u_\xi(\zeta)}f_\xi(\zeta)\Big]
\end{eqnarray}
where in terms of the two independent solutions
\begin{equation}
f'_\xi-\frac{u'_\xi}{u_\xi}f_\xi\:=\: f'^+_\xi-\frac{f'^-_\xi}{c_\xi(0)}-\frac{u'_\xi}{u_\xi}\Big(f^+_\xi-\frac{f^-_\xi}{c_\xi(0)}\Big)
\end{equation}
The obvious $\zeta$-dependence has been suppressed in the last form.
These expressions are to be used in the spectral density of the time series of the magnetic fluctuations
\begin{equation}
\big|B_\omega(x)\big|^2\ \longmapsto\ \Big|u_\xi^2(\zeta)A'_\xi(\zeta)\Big|^2 
\end{equation}
These simplify somewhat because the solutions are linearly independent. Moreover, ultimately only the observations at $x_0=0$ are available which leads to further simplification:
\begin{equation}
\big|B_\omega(0)\big|^2\longmapsto \big|u_\xi(0)\big|^2\ \Big|f_\xi^+(0)u'_\xi(0)+f'^-_\xi(0)\Big|^2
\end{equation}
{This is the relation between the power spectrum and the two independent solutions of the Schr\"odinger equation. though it can be constructed, its nonlinear character inhibits its use in the solution of the inverse problem.} Moreover, in formation of the spectrum the phase information has been lost.}

\section{Gel$'$fand-Levitan-Marchenko equation}
The last section summarised {all} the information needed for formulating the inverse problem of {turbulent magnetic fluctuations}. In this section we reduce the inverse problem to the solution of some version of Gel$'$fand-Levitan-Marchenko's equation  \citep{gelfand1951,marchenko2011}.{\footnote{{This has been done first in the solution of the inverse geomagetic induction problem \citep{weidelt1972}. A similar problem refers to the propagation of seismic p-waves in one dimension \citep{carroll1981}. This differs from both the scattering and induction problems as it is an initial value problem where a wave is injected at a particular time $t=t_0$ with amplitude $\sim\delta(t-t_0)$ at some location (for instance an earthquake) and then passes the medium. Interest is on the properties of the medium. In turbulence this would rather correspond to the injection of energy at the outer boundary of the corona, for instance, letting it evolve into turbulence when propagating outward with the solar wind. Such a problem is very different from homogeneous turbulence in a medium which we focus on, though the latter can be considered as sitting on an otherwise homogeneous stream. Still this differs from the radially expanding solar wind who is subject to adiabatic cooling. Because adiabatic expansion counteracts turbulent dissipation, solar wind  turbulence is more complicated. We might, however, note that solving the initial value problem in the time-frequency domain is possible because we deal with a convolution problem which allows for Fourier or Laplace transforming the time away and replacing it by multiplication.}} }

The scattering formulation has mathematically lucidly been presented by \citet{koelink2008}. 
The stationary Schr\"odinger equation is a second-order ordinary differential eigenvalue equation for which we have to prescribe boundary conditions in $x$.  In an infinite medium reasonable boundary conditions are that all fields and their derivatives behave regularly at infinity $x=\pm\infty$. {In evolved turbulence we do not expect that any internal modes dominate. Thus there should not be any discrete eigenvalues.} The implication is that the frequency as well as wavenumber spectra do not exhibit any emission or absorption lines. They are unstructured. This does not necessarily mean that they are power law spectra, however. It just implies that the spectra are continuous in the entire domain $[\xi,\infty)$ and in wavenumber space. In reality this might not be the case because not all frequencies or wavenumbers will be available in the data. 

\subsection{The Jost solution} 
The general solution of the Sturm-Liouville respectively Schr\"odinger equation (\ref{eq-newind}) is, in full generality, known as the Jost solution. It is obtained by understanding the Schr\"odinger equation (\ref{eq-newind}) as an inhomogeneous differential equation with inhomogeneity $q_\xi(\zeta)=V_\xi(\zeta)$ and solving it by the method of variation of integration constants. 

{The Jost solution obtained in this way (also known as the Schr\"odinger integral equation)} is composed of the two parts 
\begin{equation}\label{eq-jost-sol}
f^\pm_\xi(\zeta)=\mathrm{e}^{\pm i\xi\zeta}+\int_{-\zeta}^\zeta\mathrm{d}y\ F_\xi(\zeta,y)\ \mathrm{e}^{\pm i\xi y}
\end{equation}
with ``initial conditions" (boundary conditions) at $\zeta=\zeta_0=0$ chosen as
\begin{equation}
f^\pm_\xi(0)=1, \qquad f'^\pm_\xi(0)=u'_\xi(0)\pm \xi
\end{equation}
Clearly, the imaginary part of $\xi$ satisfies the condition that the two above solutions behave regularly at infinity. Thus, for $\zeta>0$ for instance we require that $\Im(\xi)>0$ in $f_\xi^+(\zeta)$. Note that we had normalised all quantities to their values at $\zeta=0$. This means, for instance, 
\begin{equation}
B_\xi(\zeta)\to B_\xi(\zeta)/B_\xi(0)$ and $u_\xi(\zeta)\to u_\xi(\zeta)/u_\xi(0)
\end{equation}
Obviously one has $\zeta(0)=0$. The second condition follows from  
\begin{equation}
f'_\xi(0)=u'_\xi(0)A_\xi(0)+u_\xi(0)A'_\xi(0) = u'_\xi(0)-1/c_\xi(0)
\end{equation}
respecting the normalisations,  and using the relation $c_\xi(0)=-A_\xi(0)/A'_\xi(0)$. 

{Interesting about the Jost solution is} that the conjugate time variable, the frequency $\omega$, after mapping into $\omega\longmapsto\xi$ assumes the dimension of an inverse length and appears as a wave number. {This wave number is conjugate to $\zeta$.}  

The function $F_\xi(\zeta,y)$ cannot be chosen freely. It satisfies a number of conditions which are obtained by checking the  consistency of the Jost solution {with the Schr\"odinger equation by inserting either for $f_\xi^+$ or $f_\xi^-$ into the Schr\"odinger equation}. This implies a substantial amount of simple algebra, carrying out the various differentiations of the integrals, integration by parts, and making use of the identity $(\partial_\zeta F_\xi\pm\partial_y F_\xi)|_{y=\pm \zeta}=[F_\xi(\zeta,\pm \zeta)]'$. With $\Box\ =\partial_\zeta^2 - \partial_y^2$  the d'Alembert differential operator this leads to {the following condition for the kernel $F_\xi$ of the Schr\"odinger integral equation:} 
\begin{eqnarray}
2\Big[F'_\xi(\zeta,\zeta)&-&\textstyle{\frac{1}{2}}V_\xi(\zeta)\Big]\mathrm{e}^{i\xi\zeta}+\mathrm{e}^{-i\xi\zeta}F'_\xi(\zeta,-\zeta)-\nonumber\\
&-&\int_{-\zeta}^{+\zeta}\mathrm{d}y\Big[\Box -V_\xi(\zeta)\Big]F_\xi(\zeta,y)\mathrm{e}^{i\xi y}\ =\ 0
\end{eqnarray}
{This condition must be satisfied term by term because the exponentials are independent functions of $\zeta$.} Hence each of the terms vanishes by itself. {Consequently, the kernel $F(\zeta,y)$} satisfies itself the hyperbolic (wave) equation
\begin{equation}
{\Box}\ F_\xi(\zeta,y) = V_\xi(\zeta) F_\xi(\zeta,y)
\end{equation}
It is, in addition, subject to the conditions
\begin{eqnarray}
F_\xi(\zeta,-\zeta)&=&F_\xi(0,0)=\frac{1}{2}u'_\xi(0) \\
2\ F_\xi(\zeta,\zeta) &=& u'_\xi(0)+\int_0^\zeta \mathrm{d}y\ V_\xi(y)
\end{eqnarray}
{which follow from the first two independent terms in the above expression.} Thus $F_\xi(\zeta,y)$ as solution of the hyperbolic equation is determined if $u_\xi(\zeta)$ is known. Vice versa $u_\xi(\zeta)$ is determined once $F_\xi(\zeta,y)$ is known. {The latter is what we are aiming at. Therefore, d}ifferentiating the last expression we obtain the relation {between $F_\xi$ and the unknown potential of the Schr\"odinger equation} 
\begin{equation}\label{eq-vvv}
V_\xi(\zeta)\equiv\frac{u''_\xi(\zeta)}{u_\xi(\zeta)}=-2F'_\xi(\zeta,\zeta)
\end{equation}
{This is the wanted result, because} $V_\xi(\zeta)$ is the mapping of the turbulent response function into $(\xi,\zeta)$-space. 

The task is thus to find a way to calculate $F_\xi(\zeta,y)$. This requires construction of a relation between $F_\xi(\zeta,y)$ and the data. 
This relation is provided by Gel$'$fand-Levitan-Marchenko theory. It leads to the Volterra integral equation 
\begin{equation}\label{eq-glm}
F(\zeta,y)=G(\zeta+y)+\!\!\!\!\!\!\int\limits_{-\zeta}^{+\zeta}\!\!\!\mathrm{d}\kappa\ F(\zeta,\kappa)\Big\{G(y+\kappa)+G(y-\kappa)\Big\}
\end{equation}
for $F(\zeta,y)$ which is solved for all  $|y|<\zeta$ as function of the kernel $G(\zeta)$ which itself is provided by the observational data. Once $G(\zeta)$ is given, the Gel$'$fand-Levitan-Marchenko equation can be solved by iteration. {Derivation of this integral equation can be found in various places \citep[e.g.,][]{marchenko2011}, in stringent mathematical form in \citet{koelink2008}.} 

{For completeness we construct the above equation in a form that is applicable to our problem of turbulent magnetic fluctuations. This requires constructing the relation of $G(\zeta)$ to the observations contained in $c_\xi(\zeta)$.}

\subsection{Including the observed convolution function $c_\xi(0)$}
In dealing with turbulence we are in a situation quite different from {scattering theory \citep{gelfand1951,koelink2008}, geomagnetic induction theory \citep{weidelt1972,weidelt2005,parker1980}, seismic wave propagation \citep{carroll1981}, and the application of inverse scattering theory to the solution of nonlinear equations \citep{gardner1967}.} Except for the latter, which is of mathematical interest, they all deal with bounded regions. In scattering theory some incoming waves, either quantum electrodynamic or probability waves, are transmitted by the medium and reflected from its boundaries. In induction theory the external induction field induces an electromagnetic field inside the conducting Earth up to a certain penetration depth that depends on the frequency. In seismic wave propagation a puls starts from some point emitting waves which reflect from inhomogeneity in a bounded Earth.

In turbulence no such boundaries exist primarily. We are dealing with an infinitely extended medium filled with turbulence while observations are performed at a fixed location $x_0$. In our simplified case the medium is one-dimensional. Infinite extension implies that $x_0$ can be taken as the centre of the volume. The requirement that the fields are regular throughout the turbulent volume and behave regularly at $\pm\infty$ implies that we must distinguish between the two Jost solutions (\ref{eq-jost-sol}). 

In order to derive the Gel$'$fand-Levitan-Marchenko equation we refer to the general Jost solution and its representation through the turbulent convolution function $c_\xi(\zeta)$. Since this function is not known in all space, i.e. in one dimension along the entire length of the $\zeta$-axis, we are forced to restrict its knowledge to the only observational point $\zeta=x=x_0$ which we have put to 0. With the help of 
\begin{equation}\label{eq-bound1}
f'_\xi(0)=u'_\xi(0)-1/c_\xi(0) 
\end{equation}
and using Eq. (\ref{eq-fxipm}) we have for the Jost solution
\begin{equation}\label{eq-bound2}
f_\xi(\zeta)= \frac{1}{2}\Big[f_\xi^+(\zeta)+f_\xi^-(\zeta)\Big] -\frac{1}{2\xi c_\xi(0)}\Big[f_\xi^+(\zeta)-f_\xi^-(\zeta)\Big]
\end{equation}
The data are contained in the factor in front of the second bracket.

Assume that $f^+_\xi(\zeta)$ is defined in $\zeta>0$. Then regularity at $\zeta\to\infty$ implies $\Im(\xi)>0$ in the upper complex $\xi$-plane. Similarly, $f_\xi^-(\zeta)$ for being defined in $\zeta<0$, regularity at $\zeta\to\-\infty$ implies as well $\Im(\xi)>0$. The boundary conditions at $x=x_0=0$ then require that $f_\xi^+(0)=f_\xi^-(0)=1$ and $f'^+_\xi(0)=f'^-_\xi(0)$ are continuous, which is in agreement with Eqs. (\ref{eq-bound1}) and (\ref{eq-bound2}). Thus in case of our one-dimensional infinitely extended turbulence problem the use of the Jost solution is perfectly justified. Integration with respect to $\xi$ must be performed over the upper $\xi$-half plane. 

Inserting the Schr\"odinger integral form of the two linearly independent solutions $f_\xi^\pm$ in Eq. (\ref{eq-bound2}) and rearranging into a somewhat more convenient form yields an integral equation for $F_\xi(\zeta,y)$ as function of the data $c_\xi(0)$. This equation contains integrals and is in principle the final result. In the form found it is, however, not useful yet but can be simplified. For this purpose we multiply with $\xi c_\xi(0)$, define a new data function 
\begin{equation}\label{eq-g-data}
g_\xi(0)= \textstyle{\frac{1}{2}}\big[1-\xi c_\xi(0)\big]
\end{equation}
rearrange and write the result  in a form similar to that of ordinary scattering theory also used by \citet{weidelt1972} where it was written in a version better applicable to the bounded regime. In the infinitely extended turbulence problem we have instead
\begin{eqnarray}
&&\mathrm{e}^{-i\xi\zeta}-\xi c_\xi(0)f_\xi(\zeta)= 2g_\xi(0)\cos(\xi\zeta) - \\ [0.5ex] 
&&-\int_{-\infty}^{+\infty}\!\!\!\!\!\!\!\!\mathrm{d}y\:F(\zeta,y)\:\mathrm{e}^{-i\xi y} 
+2g_\xi(0)\!\!\int_{-\infty}^{+\infty}\!\!\!\!\!\!\!\!\mathrm{d}y\:\cos(\xi y)F(\zeta,y)\nonumber
\end{eqnarray}
We may note here that the integrals have been extended over the entire $\zeta$ space because the two Jost solutions each holding in its half space contribute by smoothly adding up through the boundary conditions at $\zeta=0$. The exponential functions in the two different $\zeta$-half spaces then add up to the cosines. 

The form of this equation reminds of a chain of Fourier transforms with respect to the dummy variable $|y|<\zeta$. One thus tries  its inverse, multiplying with $\exp(i\xi y)$ and integrating along the real axis  of the variable $\xi$ over the entire upper complex $\xi$-half space. The integral {on the left part of this equation} is analytic and does not contain any poles as long, as there are no discrete eigenvalues of the Schr\"odinger equation. Since we have excluded this case in view of the presence of a continuous spectrum of turbulence, the integral is regular and will be zero. (We may note, that one could at this place also include turbulent emission or absorption lines, corresponding to the excitation of eigenmodes in the turbulence either regular eigenmodes or intermittency. Such lines would produce a number of discrete residua on the left. This makes, however, sense only if lines are clearly identified in the data.) 

With the left-hand side zero the second term on the right is already in the form of a Fourier transform. Thus its integral just yields $-F(\zeta,y)$, and it can be brought to the left {by changing its sign}. 

The integral {of the first term} on the right can be split into two integrals by resolving the cosine into exponentials. One then recognises that $g_\xi(0)$ is a Fourier transform of the data function with respect to $\zeta$. This means it is the mapping of the frequency function into the spatial domain. Its inverse is
\begin{equation}
G(\zeta)=\frac{1}{2\pi }\int_{-\infty}^{+\infty} \mathrm{d}\xi\: g_\xi(0)\:\mathrm{e}^{i\xi\zeta}
\end{equation}
This function is to be determined from the data. {It provides the link to the observations.}

Since our reference point $x_0=\zeta_0$ in homogeneous turbulence is arbitrary, there is no prescription of the particular behaviour of $G(\zeta,y)$ at $\zeta\neq 0$. We must, however, require that the data are finite at infinity, i.e. they should vanish sufficiently fast or at least are finite at infinity. For this reason $|y|<\zeta$ is required. Thus this integral becomes $G(\zeta\pm y)$. 

Finally the last term on the right is the product of Fourier transforms, i.e. {it is the transform of a convolution integral of the data function and the unknown kernel of the Schr\"odinger integral equation. By resolving it into the configuration space convolution integral},  we ultimately arrive at the Gel$'$fand-Levitan-Marchenko equation (\ref{eq-glm}) for $F(\zeta,y)$ as given above. Its solution $F(\zeta,y)$ depends on the kernel $G(\zeta\pm y)$ which is a functional of the data of observation $g_\xi(0)$. 

In fact, the Fourier transform on $g_\xi(0)$ can formally be done. The first term just produces a $\delta$-function. The second term is the derivative of the dissipative turbulent convolution function. The result is 
\begin{equation}
G(\zeta)=\pi\delta(\zeta)-{\textstyle\frac{1}{2}}c'(\zeta)
\end{equation}
with the spatial derivative of the convolution function 
\begin{equation}
c'(\zeta)=-\frac{i}{2\pi }\frac{\partial}{\partial\zeta}\int_{-\infty}^{+\infty}\mathrm{d\xi}c_\xi(0)\mathrm{e}^{i\xi\zeta}
\end{equation}
This is a complex integral whose  integration path must be established. We have $\xi=k_0=\pm \kappa\sqrt{i}$. Hence, 
\begin{equation}
\xi\ =\ \kappa\ \exp\,(\pm i\pi/4) ~~$ where $~~ \kappa=\omega/c
\end{equation}
The positive sign holds for $\zeta>0$, the negative sign for $\zeta<0$. Integration is over the upper (lower) $\xi$-half plane depending on $\zeta$ positive (negative).  It remains to insert into the Gel$'$fand-Levitan-Marchenko equation and solve for  $F(\zeta,y)$.

Solution of the Gel$'$fand-Levitan-Marchenko equation finally provides the wanted information about the unknown turbulent dissipation function $u_\xi(0)$ respectively the potential function $V_\xi$.  Ultimately Gel$'$fand-Levitan-Marchenko theory allows determining the unknown potential function from the expression (\ref{eq-vvv})
\begin{equation}
V_\xi(\zeta)=-2 F'(\zeta,\zeta)
\end{equation}
which is the key to the construction of $u_\xi(0)$ and the turbulent response function $\epsilon^T_\xi(0)$.


We have, in principle, achieved our goal: finding a method to reconstruct the electromagnetic turbulent response function in magnetic turbulent fluctuations. The theory developed here is rather complex. However, it has a large historical record in quite different context and has been given a solid mathematical fundament.

\section{Discussion}
This is the maximum that can be achieved at the time being in the inverse problem of turbulent magnetic fluctuations in plasma. As noted earlier, it requires knowledge of the turbulent convolution function $c_\xi(0)$ which enters the last integral and through it the kernel $G(\zeta\pm y)$. This function requires measurement of the magnetic \emph{and} electric fluctuations. 

Usually only the turbulent magnetic fluctuations are available though, in principle, methods could be developed to measure the electric fluctuation field by injecting dilute ion beams into the plasma and monitoring their return fluxes which provide direct information about the low-frequency electric fluctuations. Such measurements have occasionally been performed using electrons but are polluted by the enormous sensitivity of electrons to the presence of electric and magnetic fluctuation fields. They also suffer from the difficulty of identification of the injected from ambient electrons. 

Another more promising possibility is the injection of low-energy ion beams, to measure their distribution function and to calculate from it the fluctuations of the velocity field. 

In the absence of either of these one cannot proceed further. Magnetic field observations alone are insufficient. They cover only half of the information stored in the electromagnetic field.  

It is easy to see that without the independent determination of the turbulent convolution (response) function $c_\xi(0)$ one cannot proceed. It can be written as a differential equation for the vector potential respectively electric field component
\begin{equation}
A'_\xi(\zeta)+A_\xi(\zeta)/c_\xi(0)=0
\end{equation}
whose solution at fixed frequency $\xi$ 
\begin{equation}
A_\xi(\zeta)=A_\xi(0)\exp[-\zeta/c_\xi(0)] 
\end{equation}
shows that $c_\xi(0)$ is the typical scale of variation of the electric field respectively the vector potential in $\zeta$. The observations provide instead the \emph{frequency spectrum} of the magnetic field $B_\xi(0)$ which is the spatial derivative in $x$ of the vector potential at location $x=\zeta=0$. This derivative contains the unknown function $u_\xi(\zeta)$
\begin{equation}
B_\xi(0)=\partial_x A_\xi(0)= u^2_\xi(0)A'_\xi(0)
\end{equation}
When using  $A_\xi (\zeta)= u_\xi(\zeta) f(\zeta)$ this becomes
\begin{equation}
u^2_\xi(0)A'_\xi(0)= u_\xi(0)f'_\xi(0)-u'_\xi(0)f_\xi(0)
\end{equation}
The initial or boundary  conditions $f_\xi(0)=1$ and (\ref{eq-bound1}) imply that
\begin{equation}
B_\xi(0)= {\textstyle\frac{1}{2}}{[u_\xi(0)-1]^2}'-u_\xi(0)/c_\xi(0)
\end{equation}
which still contains the unknown function $u_\xi(0)$. 

This simply expresses the above noted obvious fact that reduction to magnetic measurements alone, lacking the electric field or otherwise the velocity field, {implies loss of one half of the electromagnetic information which is needed in solving the inverse scattering problem}. This resembles the inverse scattering case where without knowledge of the reflection and transmission coefficients which couple the incoming and outgoing waves, no solution exists. Hence, in solving the inverse problem of turbulent magnetic fluctuations \emph{knowledge of either the electric field or velocity fluctuations in addition to the magnetic fluctuations is obligatory}.

With the reduction of the inverse problem of turbulent magnetic fluctuations to the Gel$'$fand-Levitan-Marchenko equation the formal problem of inversion of the magnetic fluctuations in a turbulent plasma has been solved. It has been reduced to the determination of the dissipative convolution function from observations of the fluctuations of the electromagnetic fields at observation point $x_0=0$. 

In practice the full solution of the inverse problem which aims at the determination of the dissipative response function $\epsilon^T_\xi$ requires providing the data in treatable form, solving the integral equation, and afterwards calculating the response function. These three tasks are still open for handling. {Thus reduction of the inverse problem to the Gel$'$fand-Levitan-Marchenko  integral equation is just an important and necessary though only an intermediate and not yet the ultimate sufficient step.}


The form of the convolution function is not known a priori. Its spatial dependence is not required, however. Necessary is just its temporal spectrum i.e. its Fourier transform with respect to $\xi$. Though this is not known, from some analogy to the magnetic frequency spectra and the models of magnetic turbulence one may expect that the convolution function has similar formal properties. 

We already noted that it is not expected that the turbulence contains emission or absorption lines corresponding to eigenmodes of the Schr\"odinger equation. This would imply the presence of distinct plasma waves or turbulent energy losses at some particular frequency as might be present in non-homogeneous plasmas such as like near a shock wave in a restricted region of space. Examples are the narrow upstream and downstream regions of shocks in the solar wind, the foreshock and magnetosheath regions where turbulence prevails while distinct plasma modes are excited by some energy source related to the shock. Moreover, if the turbulence is not fully developed, intermittency might play a role leading to additional structure in the dissipative response function. 

{These problems are all very interesting and important. They, however, as noted several times, in order to be included into the inverse problem require precise measurements of both types of electromagnetic fields, its magnetic and electric components. Formally they introduce discrete eigenvalues leading to poles in the complex $\zeta$ plane which generate residues. These should appear as additional terms in the Gel$'$fand-Levitan-Marchenko equation. They also cause modification of the data function $g_\xi(0)$ which enters the kernel of the integral equation.}

\subsection{Power law fluctuation spectra}
In the absence {of discrete fluctuation modes}, we may try an unstructured power law distribution of the dissipative convolution function of the kind
\begin{equation}
c_\xi(0)= a_0 \xi^{-\alpha} $ ~~for~ ~$ \xi_0 < \xi <\xi_d
\end{equation}
with some dimensional factor $a_0$. The limitations of the frequency range are arbitrarily assumed, with $\xi_0$ some low frequency cut-off of the spectral power law range, and $\xi_d$ some high-frequency cut-off which possibly can be related to the onset of strong dissipation which breaks the power law. Such a power law may be justified by assuming that both the electric and magnetic fluctuation  fields obey unstructured power law spectra in frequency space, 
$E_\xi(0)\sim \xi^{-\alpha_E}, B_\xi(0)\sim \xi^{-\alpha_B}$. Since $c_\xi(0)=-E_\xi(0)/\xi^2cB_\xi(0)$, the ratio of the two power laws yields another power law $\alpha=\alpha_E-\alpha_B+2$ with the various constant factors and normalisations combined into the constant $a_0$. {It should be stressed that these power laws are not power laws of spectral energy densities inferred in turbulence theory; rather they are simply the frequency spectra of fluctuations if they exist in this form}.{\footnote{{Reformulation of the convolution function in terms of turbulent spectral energy densities would require a complete reformulation of the inverse problem which we do not intend in this work. Such a reformulation suppresses the use of Jost functions which are the solutions of the Schr\"odinger equation for the mapped fields, not for their spectral energy densities. Presumably this inhibits reference to the Gel$'$fand-Levitan-Marchenko theory.}}} 

Actually, the Fourier transforms map the electric and magnetic fluctuation fields from time into frequency space. The mapped field spectra necessarily possess some phases $\phi_\xi^{E,B}(0)$. The convolution function is thus itself a function of the phase differences between the electric and magnetic Fourier spectra. In homogeneous turbulence their phases and thus also their phase difference can be assumed to be randomly distributed. They can, in principle, be averaged out in this case, just contributing to the factor that multiplies the assumed power law. This assumption is followed below.

{The assumption of a power law of the turbulent convolution function allows to  write
\begin{equation}
g_\xi(0)={\textstyle\frac{1}{2}}\Big[1-a_0\ (\xi-\xi_0)^{-\alpha}\Big] $~~ with ~~$\Re\,{\alpha} > 0
\end{equation}
which inserted into the inverse transform of $g_\xi$ yields the kernel function 
\begin{equation}\label{eq-kernel-g}
G(\zeta) =\pi\delta(\zeta)-\frac{a_0}{4\pi i}\frac{\partial}{\partial\zeta} \int_{0}^{\xi_d} \mathrm{d}\xi (\xi-\xi_0)^{-\alpha}\mathrm{e}^{i\xi\zeta}
\end{equation}
Since $\xi=\kappa\mathrm{e}^{\pm i\pi/4}$ is a complex wavenumber, we have $\mathrm{d}\xi=\mathrm{d}\kappa\ \mathrm{e}^{\pm i\pi/4}$. The integral becomes a  contour integral in the complex $\xi$ plane limited by $\xi_d=\kappa_d$, a real frequency respectively wavenumber corresponding to the power law range of  the observations in frequency. Integration is thus along the real axis between zero and these limits, closed by a large circle over the upper (lower) half of the complex $\xi$-plane up to an angle $\pm\pi/4$ and returning on the ray along this angle to zero, depending on the sign of $\zeta$. One may note that there is a singularity on the real axis at $\xi=\xi_0$ which is of fractional order $\alpha$ giving rise to fractional Riemann branches which is seen when writing $(\xi-\xi_0)^{-\alpha}=\exp[-\alpha\:\ln|\xi-\xi_0|-\alpha i\theta]$. However, for an expected value $\alpha>1$ the complex integration contour lies completely in the principal branch making integration around this pole possible without caring for the branch cuts. It gives rise to a residuum $2\pi i\:\exp(i\kappa_0\zeta)$. This would be the total value of the total contour integral around the integration contour. What is wanted, is the principal value along the real axis across the singularity at the point $\xi_0$. This could be determined when the remaining parts of the contour integral are found. Calculation of the remaining circular section at $\xi_d$, and the ray at angle $\pi/4$ back to the origin, is however difficult and cannot be given in closed analytical form even when shifting the upper limit $\xi_d \to \infty$. The problem is the integral along the ray. Thus determination of the integral in Eq. (\ref{eq-kernel-g}), i.e. the principal value of the integral, is not easily done.}

{If we, for simplicity, assume that the power law spectrum extends over several orders of magnitude in $\xi$,  we can put $\kappa_0\ll\kappa_d$. Then the limits of integration can be pushed to their extremes $0$ and $\infty$. We don't know the value of $\alpha$. One expects $0<\alpha<3$ to be a fraction corresponding to some root. Turbulent spectra of both the electric and magnetic fields are smooth. Consequently, the power law of the turbulent convolution function it smooth as well, and the integral does not contain any poles other than $\xi_0$. Shifting the origin into $\xi_0$ yields the residual factor $\exp(i\kappa_0\zeta)$, and one can solve the remaining integral by the method of steepest descent. }

{We write the integral with respect to $\kappa$ and the integrand as an exponential $\exp \psi(\kappa,\zeta)$ with
\begin{equation}
\psi_\pm(\kappa,\zeta) = \mp\frac{i\pi\alpha}{4}-\alpha\ln\kappa+i\kappa\zeta\mathrm{e}^{\pm i\pi/4}
\end{equation}
Its first derivative with respect to $\kappa$ put to zero yields the fixed point 
\begin{equation}
\bar\kappa_\pm= (\alpha/\zeta)\mathrm{e}^{\mp3i\pi/4}
\end{equation}
The second derivative taken at the fixed point is 
\begin{equation}
\psi''_\pm(\bar\kappa,\zeta)=\alpha/\bar\kappa_\pm^2=\mp i\zeta^2/\alpha 
\end{equation}
with both $\zeta$ and $\alpha$ real. The exponent can now be expanded around the fixed point up to second order, which yields for the integral
\begin{equation}
\int\limits_{\xi_0}^{\xi_d}\mathrm{d}\xi\xi^{-\alpha}\mathrm{e}^{i\xi\zeta}\ =\ \mathrm{e}^{\psi_\pm(\bar\kappa,\zeta)}\int\limits_{\kappa_0}^{\kappa_d}\mathrm{d}\kappa\mathrm{e}^{\frac{1}{2}\psi''_\pm(\bar\kappa,\zeta)(\kappa-\bar\kappa)^2}
\end{equation}
The formal solution of this Gaussian integral is the difference between two error functions
\begin{eqnarray}
\int\limits_{\xi_0}^{\xi_d}\mathrm{d}\xi\xi^{-\alpha}\mathrm{e}^{i\xi\zeta}\ &=&\ \sqrt{\frac{\pi/2}{\psi''_\pm(\bar\kappa,\zeta)}}\bigg\{\mathrm{erf}\bigg[\sqrt{{\textstyle\frac{1}{2}}\psi''_\pm(\bar\kappa,\zeta)}(\kappa_d-\bar\kappa_\pm)\bigg]\nonumber\\
&-&\mathrm{erf}\bigg[\sqrt{{\textstyle\frac{1}{2}}\psi''_\pm(\bar\kappa,\zeta)}(\kappa_0-\bar\kappa_\pm)\bigg]\bigg\}
\end{eqnarray}
The error functions are functions of complex arguments whose convergence properties with respect to the variable $\zeta$ must be taken into account when inserting into the kernel function $G(\zeta)$, in particular when the boundaries are allowed to take their extremal values $\kappa_0=0,\kappa_d=\infty$. (One may note that turbulent spectra consisting of several piecewise parts obeying different power laws which smoothly connect via spectral break points can be included. In this case the above sum of error functions multiplies by the number of such ranges, each with a different power $\alpha$, i.e. different $\bar\kappa_\pm$.)  It turns moreover out that the dependence of the kernel on the power $\beta$ of the power law is very complicated. One therefore expects that the solution of the inverse problem, i.e. the solution of the Gel$'$fand-Levitan-Marchenko equation will not be possible to be constructed analytically, even in the simple case of power law spectra. In the limit of an extended spectrum the integral becomes 
\begin{equation}
\int\limits_{0}^{\infty}\mathrm{d}\xi\ \xi^{-\alpha}\mathrm{e}^{i\xi\zeta}\ =\ \sqrt{\frac{\pi/2}{\psi''_\pm(\bar\kappa,\zeta)}}\ =\ \sqrt{\frac{\pi\alpha}{2}}\frac{\mathrm{e}^{\pm i\pi/4}}{\zeta}
\end{equation}
This has to be multiplied by $\exp\psi_\pm(\bar\kappa,\zeta)$, the factor in front of the integral, which is another complicated function 
\begin{equation}
\mathrm{e}^{\psi_\pm(\bar\kappa,\zeta)}=\bigg(\frac{\zeta}{\alpha}\bigg)^\alpha\exp\bigg[\frac{i\pi}{4}(\alpha\pm\alpha)+\frac{(i\mp 1)\alpha^2}{2\sqrt{2}\: \zeta}\bigg]
\end{equation}
Combining all these expressions, the data function $G(\zeta)$ is determined  for use in the Gel$'$fand-Levitan-Marchenko equation (\ref{eq-glm}). As before the signs $\pm$ apply to the signs of the argument in $\zeta$ which map to the arguments in $G(\zeta\pm y)$ when accounting for the integration with respect to $y$. This by itself becomes a major analytical and numerical effort. 
We leave the solution of the Gel$'$fand-Levitan-Marchenko equation in the particular case of a power-law  turbulent-convolution function for a separate investigation. }

{Use of a power law spectrum for the temporal fluctuations is justified by observations which provide the turbulent electromagnetic fluctuations. The relation to any kind of Kolmogorov spectra \citep{kolmogorov1941a} and its apparent observation in space plasmas \citep{goldstein1995,zhou2004} is not clear however. The assumption of a power law convolution function $c_\xi$ has been purely artificial. Observations provide spectral energy densities while we used the fluctuation spectra. There is clearly a relation between the two (see before); the inverse turbulence problem is, however, not defined in terms of the spectral energy density. }

{Observations in the solar wind suggest that power laws are realised in the spectral energy densities only over a number of ranges of very limited extension in frequency. Observed spectra contain break points which connect spectral ranges exhibiting different power laws. They also contain more or less well expressed energy injection as well as dissipation ranges of various shapes ranging from exponential decay to exponentials of more complicated arguments or even algebraic decays. Hence, assuming a simple power law in the turbulent convolution function and extending the range of integration over the entire frequency interval from zero frequency to infinity somehow violates the observational input. The solution for a power law turbulent convolution function can therefore serve only as an example and has little to do with reality.}

{In fact, if anyone wants to apply the inverse procedure to power spectral energy densities he must refer to Poynting's theorem in electrodynamics, i.e. use the heat transfer equation for the electromagnetic field. This equation is nonlinear containing the product $\vec{j\cdot E}$ and the divergence of the Poynting vector. It seems improbable that this equation can be easily transformed into linear Schr\"odinger-like form and application of the Gel$'$fand-Levitan theorem as this is reserved for the linear Schr\"odinger equation only. One may conclude that though the inverse problem of turbulent fluctuations can well be treated by this approach, the inverse problem of spectral power densities inhibits such an approach for the above mentioned reasons. Moreover, in any case it will be necessary to include measurements of the electric power spectrum respectively fluctuations because only the full electromagnetic field contains the full electromagnetic information about the dynamics.}

Doing justice to the observations necessarily implies a numerical treatment of the inverse problem of magnetic turbulence. It moreover requires, in addition, the measurement of both the electric and magnetic fluctuation time series and subsequent determination of their spectral equivalents.  

\subsection{Conclusions}
The possibility of an application of the inverse problem of scattering to turbulence and fluctuations has not been obvious. It required the transformation of the electromagnetic turbulent fluctuation problem into the Sturm-Liouville-Schr\"odinger form. This is an interesting turn that might bring a new view on magnetic fluctuation and turbulence theory, possibly opening a path to infer the properties under which magnetic fluctuations in a broad spectral range in plasma develop. 

Solving the new integral equation for a given data set is actually not an easy task. Though the solution of the Gel$'$fand-Levitan-Marchenko equation should not provide unsurmountable hurdles, it is not particularly simple. The way to solve it starts from the assumption of any reasonable initial solution for the unknown function $F(\zeta,y)$ and to iterate. In many cases  an approach of this kind leads to fractional chains which, depending on the first intelligent guess, should rapidly converge such that not many steps should be necessary to perform. Of course the  result and procedure depend on the resolution of the time series of the magnetic and electric fluctuations and their fluctuation spectra. 

The general method is not restricted to a simple power law in the fluctuations just covering the limited inertial range as used in the above last example. It depends on the precision of the time resolution of the observations and the related spectral representation of the turbulent magnetic fluctuations. This spectrum may well extend from the lowest injection frequencies up to deep into the dissipation range \citep{alexandrova2009,sahraoui2009,sahraoui2013}. It thus should reproduce the turbulent response function over the accessible range of frequencies merely excluding the genuinely kinetic regime at the highest frequencies, i.e. the regime where turbulent magnetic fluctuations do not provide any direct information about the turbulent electrostatic contributions generated by the plasma dynamics.  These have been excluded by the assumption of purely transverse magnetic fluctuations and turbulence. 

Otherwise all magnetically active dynamical contributions to the evolution of turbulence will contribute and are formally included in the theory in the definition of the response function. Moreover, since the theory is based on temporal spectra as ingredients of the observational function $G(\zeta)$, no reference to the Taylor hypothesis is required. This also implies that it is not required to map the spectrum into the wavenumber range. The inversion theory does include this transformation anyway as we have noted above by attributing $\xi$ as the Fourier-conjugate to the spatial coordinate $\zeta$. The method does, in this way, provide information about the way the full turbulent spectrum is generated. This is the physically interesting part of the theory. 

The only restrictions on the validity of the Gel$'$fand-Levitan-Marchenko approach in its form of making it available to magnetic turbulent fluctuations are the simplifying assumptions made by us of one-dimensionality, homogeneity of turbulence, restriction to non-expanding turbulence, the uncertainty of observations and thus the incompleteness of coverage of the frequency domain. Some of these assumptions may prevent application to fast expanding plasma streams like the solar wind where the observations are performed in one particular spatial point which is about fixed to space and not to the stream thus violating one of our assumptions.
  
Though the relation between the observations in our one-dimensional approach and the inverse theory is striking, it is obvious from the last expression that even the solution of the Fourier integral, the input to the kernel of the Gel$'$fand-Levitan-Marchenko equation, cannot be provided in a sufficiently simple form that would allow for an analytical solution. This does not prevent the application of the theory; it only suggests that any application must necessarily not only refer to numerical work in establishing the power spectrum of turbulence by observations, it also requires a subsequent numerical treatment of the inverse problem. Whether or not this will be advantageous in investigating turbulence is hard to estimate. The effort in formulating and solving the inverse problem is large. Its outcome is the maximum available information about the turbulent response function at maximum effect of the fluctuating fields on it. This function will subsequently have to be interpreted physically in view of the conditions under which turbulent magnetic fluctuations have evolved. An reformulation of this theory to the inclusion of the turbulent power spectral densities on which current investigations of magnetic turbulence rely is, however, currently not in site. Its formulation would require use of Poynting's theorem in turbulence and an attempt to transform it into Schr\"odinger form which, probably, cannot be done.  

\section*{Appendix: Derivation of Equation (\ref{eq-newind})}
Here, for completeness and since the algebra is nontrivial, we provide the transformation of the induction equation 
\begin{equation}
A_{xx}=\xi^2\epsilon^T_\xi(x) A\quad\mathrm{with}\quad A_x\equiv\mathrm{d}A/\mathrm{d}x
\end{equation}
into the Schr\"odinger-like form
\begin{equation}
f''(\zeta)=\Big[\xi^2+V(\zeta)\Big]f(\zeta),\ \ \ \zeta=\int_0^x\mathrm{d}t\sqrt{\epsilon^T_\xi(t)}
\end{equation}
Define $u(\zeta)\ :\ =\sqrt[4]{\epsilon^T_\xi(\zeta)}$ and $f(\zeta)=u(\zeta)A(\zeta)$. Then one has $\mathrm{d}\zeta/\mathrm{d}x=u^2(\zeta)$.  Divide the above original induction equation by $\epsilon^T_\xi=u^4$ and use\ $' : \ =\mathrm{d}/\mathrm{d}\zeta$ to obtain
\begin{eqnarray}
f'&=&u'A+uA' \nonumber\\
f''&=& u''A+2u'A'+uA''\nonumber
\end{eqnarray}
Now, $A'=A_x/u^2$. Thus
\begin{equation}
f''=u''A + \Big\{2(u'/u^2)A_x +uA''\Big\} 
\end{equation}
One must express $uA''$ in order to recover $A_{xx}$. This is done by calculation as follows:
\begin{equation}
A'' =\bigg(\frac{A_x}{u^2}\bigg)'=\frac{A_{xx}}{u^4}-\frac{2u'}{u^3}A_x
\end{equation}
Multiply by $u$ and use in the first equation and in the curly brackets of the last above expressions for $A_{xx}$ and  $f''$, rearrange and define $V(\zeta)=u''(\zeta)/u(\zeta)$. This produces the wanted Schr\"odinger form of the original induction equation. It is clear that this form cannot be trivially obtained. 

In order to finally, after solving the inverse problem, recover the spatial coordinate $x$ one has to perform the integral
\begin{equation}
x(\zeta) = \int_0^\zeta\mathrm{d}t/u^2(t)
\end{equation}
This can be done because the solution of the Gel$'$fand-Levitan inverse problem produces $u(\zeta)$ from the observational data and, moreover, $\Re\,{u}^2(\zeta)> 0$ for any realistic case.

%
%





\end{document}